\documentclass [10pt]{article}
\usepackage{amsmath}
\usepackage[dvips]{graphicx}

\begin{document}

\begin{center}
\Large{Variation in the Earth's figure axis owing to geophysical cause,\\
minutely determined from polar motion}

~\\
\normalsize{August, 2014\\
Yoshio Kubo}\footnote{5-8-3 Higashi-gotanda, Shunagawa-ku, Tokyo, 141-0022 Japan.  (kuboy@sakura.to)}
\end{center}

\begin{center}
\textbf{Abstract}
\end{center}

The variation in the figure axis  in the Earth owing to the geophysical causes directly reflects the physical state of the Earth
and therefore, it is important for understanding the Earth.
Of the variations in the figure axis arising from different causes, the variation owing to the Earth's geophysical causes
is determined from polar motion data.
Equations that give the relation between the figure axis and rotational axis are derived and
the motion of the figure pole with respect to the Earth's surface or the figure axis in the Earth is determined by using these equations.
Estimation of error suggests that the accuracy of the determination for the position of the figure pole
at each moment is sufficiently good and that even a variation that occurs during a short duration of time, such as one or two days, is significant.
The determined position of the figure pole exhibits a rapid and complicated motion in a short duration.
On the other hand, a long-range average of the pole shows stable annual and semi-annual variations, but it is quite independent of the Chandler wobble.

~ 

\section{Introduction}

Polar motion is a phenomenon of the variation in the position of the rotational pole on the Earth's surface in the North Pole region.

Polar motion shows an irregular feature that reflects the physical state of the deformable Earth.
For that reason, numerous analyses have been published on polar motion, especially on its periodic parts.
Some studies, for example, Kimura (1917), Guinot (1970) and H$\ddot{\textrm{o}}$phner (2002), 
aim at only analyzing the pattern of the polar motion itself, such as the variations in its amplitude and phase.
On the other hand, some other studies intend to explain the variation in the polar motion by geophysical causes,
such as the atmosphere (Aoyama and Naito, 2001), ocean (Dickman,1993;  Gross, 2000) geomagnetic jerks (Gibert and Mou$\ddot{\textrm{e}}$l, 2008) and so on.

However, the physical condition of the deformable Earth first brings about a change in the mass distribution in the Earth.
Then it deforms the shape of the Earth including the position of the figure axis,
which is the axis with the largest moments of inertia of the Earth's three principal axes
or the symmetric axis if the Earth is regarded as spheroidal.
The variation in the figure axis in the Earth then affects the rotational axis or the rotational pole and
the change in its position is the polar motion.
Therefore, it is the variation in the figure axis or the figure pole but not the polar motion itself
that should be linked to the physical state of the Earth.

As the motion of the figure pole determines the polar motion, the position and motion of the figure pole at each moment
can be obtained from the polar motion data as stated below.
However, studies on obtaining the position and variation of the figure pole (Wilson (1985) is an example.) are few, 
in spite of the fact that data on polar motion are monitored and published regularly and
consequently we possess a satisfactory data set of it.

Polar motion is the motion of the rotational pole on the Earth's surface.
Therefore, if we know the relation between the rotational pole and the figure pole,
we can determine the position of the figure pole on the Earth's surface.

The motion of the rotational pole relative to the figure pole, or that of the rotational axis to the figure axis 
has been studied thoroughly by methods of both geophysics and dynamical astronomy.

In the case of a rigid Earth, the figure axis is fixed in it and, if a spheroidal Earth is assumed, the polar motion
is a perfect circular motion around the figure pole and it is called the Euler motion.
The angular speed of the Euler motion is given by $[(C - A)/A]\omega$, in which $C$ and $A$ are the moments of inertia along the symmetrical axis
and an axis on the equatorial plane, respectively, and $\omega$ is the rotational speed of the Earth.  
The corresponding period is about 304 days with the observed values for the ratio $(C - A)/A$ and $\omega$.
  
In the actual Earth, the period of the Euler motion suffers a change to that of the Chandler wobble (about 437 days)
mainly owing to the deformation caused by the centrifugal force brought about by the rotation.
In addition, the center and radius of the Chandler wobble change at all times,
resulting in a complicated polar motion.
However, even in the actual deformable Earth, the rotational pole instantaneously makes a circular motion around the figure pole at that instant.
Since the position of the figure pole changes irregularly, the polar motion is not simple.

There are three kinds of variations in the figure axis.  
The first variation is caused by the elastic deformation of the Earth produced by the attraction from the outer bodies:
the Moon and the Sun.
The second variation is caused by the deformation owing to the centrifugal force brought about by the rotation.
The last variation is owing to the deformation of the Earth caused by the Earth itself including the ocean and air.

The first two variations can be predicted by using well-determined motions of the celestial bodies and the Earth
and by supposing an Earth model with a suitable parameter for the elasticity.
On the other hand, the third variation cannot be predicted exactly because it is brought about by irregular motion
of the mass in the solid Earth, the ocean, and the air, that is, by geophysical causes.

However, this variation can be determined by using polar motion data.
Further, as the accuracy of the observations for polar motion is very high, the variation in the figure axis 
owing to geophysical causes is expected to be determined  with a high accuracy as well.

The present study intends to determine the variation in the Earth's figure axis brought about by the Earth's deformation 
owing to causes other than the attraction from outer bodies and the centrifugal force arising from rotation.

\section{Definitions of the figure and the rotational axes used in the present study}

The polar motion in the rotation of the Earth is expressed as the motion of the rotational pole referred to a coordinate system set on the Earth's surface
in the North Pole area.
We separate this motion into two parts:  First, the motion of the Earth's figure pole in the said coordinate system
and the second is the motion of the rotational pole relative to the moving figure pole.
In the rotation of a rigid Earth, the former motion does not exist but only the latter does.

It is important first to give exact definitions for both the figure axis and the rotational axis, or the respective poles, used in the present study.

As for the rotational pole in the present study, we adopt the Celestial Intermediate Pole (CIP),
which is defined clearly through a resolution of the International Astronomical Union (IAU, 2000).
CIP (called the Celestial Ephemeris Pole (CEP) before) defines the fundamental coordinate system of date on the celestial sphere.
Precession and nutation in the astronomical ephemeris are defined as the secular and periodic motions of CIP, respectively,
with respect to an inertial system.  
CIP is the pole or the axis around which the Earth actually rotates
but it is different from the mathematical definition of the rotational axis by less than $0.01''$ (Seidelmann, 1982; IERS, 2004).

The motion of the rotational pole that is defined by the CIP does not contain a diurnal component on the celestial sphere nor with respect to the Earth's surface.
As a result, polar motion as defined by this rotational pole does not have a diurnal component.
More importantly, the coordinates of the pole in the polar motion published by the International Earth Rotation Service (IERS),
which we use as the basic data in the present study, are also based on this pole.
So, we adopt the same definition of the rotational pole that is adopted by the IERS.

Next, we present the definition of the figure axis or the figure pole used in the present study.
The figure axis of the Earth, which varies at every instant, is the axis with the largest moment of inertia at this instant.
It varies with the change in the Earth's mass distribution owing to both the inner and outer causes as mentioned above.
Of them, the variation owing to the latter cause has to be first removed.

In fact, the instantaneous circular motion of the rotational pole does not occur around the figure pole brought about
by the outer forces, but it is around the averaged figure pole with regard to the deformation owing to the outer force
as shown in Kubo (2009) as well as in Moritz and Mueller (1987).
Of these two references, the rotational pole in Kubo (2009) is the same as that defined above;
however, the definition is not very strict for the rotational pole in Moritz and Mueller (1987).
Further, in the latter, the figure pole without the outer force and the rotation is supposed to be fixed.

It should be noticed that the motion of the pole brought about by the external cause is nearly circular and diurnal on the Earth's surface
and it gives no information as for the changing physical state of the Earth at that moment.
Therefore, we are not interested in it in the present study although the radius of this roughly circular motion amounts to several tens of meters or about $2''$ at the maximum.
In addition, if necessary, it can be calculated as exactly as required.

On the other hand, we must pay special attention to the effect of the centrifugal force on the figure axis.
The figure axis is considerably shifted by the centrifugal force owing to the rotation 
from the position that the axis would take if the Earth were not rotating.

This shift in the figure axis has its cause from within the Earth but not from outside.
In addition, it has the same period as that of the Chandler wobble, not nearly diurnal, on the Earth's surface.
The rotational pole actually makes a circular motion around this figure pole that is shifted by the centrifugal force.
However, the rotational pole makes an instantaneous circular motion virtually around the original figure pole that is not shifted
by the centrifugal force because of a mechanism that is explained in the next section.
Therefore, the shift in the figure pole brought about by the rotation is also excluded from our consideration.
In addition, the amount of the shift by this cause can be calculated exactly based on dynamics.
We are interested only in the variation in the figure axis by the causes originating in the Earth
that cannot be predicted in a dynamical sense.

To sum up, the figure axis treated in the present study is that which the Earth would have had if there were neither outer forces nor rotation.

\section{Equations describing the rotational pole relative to the figure pole}

Now, we consider the relation between the rotational pole and the figure pole as defined in the previous section through rigid dynamics,
by following Kubo (1991) and Kubo (2009).

As mentioned in Sect. 1, in the free rotation of a rigid Earth, the rotational pole rotates around
the figure pole with an angular speed of $\displaystyle \frac{C - A}{A} \omega$,
where $\omega$ is the sidereal angular speed of the Earth's rotation (Kinoshita, 1977).

A similar circular motion of the rotational pole also occurs in the deformable Earth.
We suppose that the Earth is deformed by any cause and consider a rigid Earth with the same values of $A$ and $C$ as in the deformed Earth  
and with the figure axis coincident with that which is shifted by the deformation, that is, a rigid Earth which is equivalent to the deformed Earth.
Then, the rotational pole of the deformed Earth behaves in the same way as in the equivalent rigid Earth at each instant (Kubo, 2009).
In the discussion in this section, we completely neglect the deformation by the force from the outer bodies for the reason mentioned in the previous section.

\begin{figure}
\centering
\includegraphics[width = 5cm, clip]{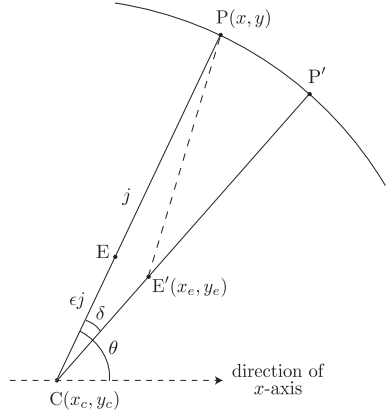}
\caption{Earth's surface near the North Pole.  C is the figure pole without the effect of the rotation
and E or E$'$ is the figure pole affected by it, each of them at some moment.
P is the rotational axis at the same moment and P$'$ is that at some time duration earlier.
The origin of the xy-coordinate system is at the IRP (International Reference Pole)
and the x- and  y-axes are in the directions of the Greenwich and the $90^{\circ}$ east longitude, respectively.}
\label{NPole}
\end{figure}%

We consider first that the Earth has elasticity but not plasticity.
Fig. \ref{NPole} shows the Earth's surface near the North Pole.  Let C and P be the figure pole and the rotational pole at some instant 
as defined in the previous section (Note that C moves with time as well.), and let $\overline{\textrm{C} \textrm{P}} = j$. 
The bulge by the centrifugal force owing to rotation is produced at the equatorial zone of the rotational axis.
According to Kubo (1991), the figure axis which is produced as the compound of the original figure without the effect of rotation
and the bulge produced owing to the rotation is located at the point E on the line CP and at the distance $\epsilon j$ from C, with
\begin{equation}
\epsilon = \frac{3\mu C}{C - A},
\end{equation}
where $\mu$ is given by
\begin{equation}
\mu = \frac{ka_e^2 {\omega}^2}{9GC} = 0.00116k,
\end{equation}
$k$ being the Love number, $a_e$ the equatorial radius of the Earth, and $G$ the universal gravitational constant.

Then, as for the behavior of the rotational pole toward the shifted figure pole at that instant,
P moves in such a way as to rotate around E with the angular speed
\begin{equation}
\Omega = \frac{C - A}{A}\omega. 
\end{equation}
It should be noticed that $A$ and $C$ in this equation are those for the actual Earth shape including the centrifugal effect,
that is, they are the observed values for the moments of inertia.  Therefore, the value of $\Omega $ is $(1/304)\omega $ and the corresponding period is
about 304 sidereal days.  This is equal to the angular speed of the Euler motion in the rotation of the equivalent rigid Earth.

Because P makes such a motion at each instant, it rotates around C with an angular speed
\begin{equation}
\Omega ' = (1 - \epsilon)\Omega = \frac{(1-3\mu)C - A}{A}\omega. 
\end{equation}
Supposing $k = 0.29$, this angular speed corresponds to the period of about 440 days.
Thus, the period of the Euler motion is prolonged owing to the elasticity of the Earth,
being coincident with the expression in Kubo (1991), which was derived in a more analytical way.

Next, we consider the Earth that has, besides elasticity, anelasticity.
We treat this condition by introducing a model such that
the bulge by the centrifugal force at a moment is produced not around P, which is the rotational axis at the moment,
but around P$'$, the rotational axis some time interval ago (Fig. \ref{NPole}).

Let $\angle \textrm{P}\textrm{C}\textrm{P}' = \delta $.  Then, the actual figure axis at the moment comes to the point E$'$ on the line CP$'$ and
at a distance $\epsilon j$ from C. 
We consider that C does not move and $j$ does not change during the time interval $\delta /\Omega' $,
which corresponds to the time for the rotational pole to move from P$'$ to P (time length probably of one day or so).
This results in P moving so as to draw a circle around E$'$ with the radius $\overline{\textrm{E}'\textrm{P}}$,
which is equal to $(1 - \epsilon)j$ with a sufficient approximation and an angular speed of $\Omega $.

This situation is expressed by the following equations:
We consider a coordinate system on the Earth's surfacein the North Pole area.  Although this coordinate system is arbitrary, it may be convenient to choose
the International Reference Pole (IRP), which can be regarded as practically the same as the Conventional International Origin (CIO), as the origin
and take the x-axis in the longitude of Greenwich.  Then, the y-axis is in the $90^{\circ}$ east longitude.

Let the coordinates of P, C and E$'$ be $(x, y), (x_c, y_c)$ and $(x_e, y_e)$, respectively.  Then, the velocity of P $(\dot{x}, \dot{y})$ is written as
\begin{eqnarray}
\dot{x} & = & -(y - y_e)\Omega, \nonumber \\
\dot{y} & = & (x - x_e)\Omega .
\label{prim}
\end{eqnarray}
We then express $(x_e, y_e)$ in terms of $(x_c, y_c)$ in these equations.  Let the angle of the direction of the line CP from the x-axis be $\theta$.  Then, 
\begin{eqnarray}
x_e & = & x_c + \epsilon j \cos(\theta - \delta) \nonumber \\
\displaystyle
    & = & x_c + \epsilon j \left( \frac{x - x_c}{j} \cos \delta + \frac{y-y_c}{j}\sin \delta \right) \nonumber \\
    & \cong & x_c + \epsilon[(x- x_c) + (y - y_c)\delta].
\end{eqnarray}
Therefore, we have, with sufficient accuracy
\begin{eqnarray}
x - x_e & = & x - x_c - \epsilon[(x - x_c) + (y - y_c)\delta] \nonumber \\
        & = & (1 - \epsilon)(x - x_c) - \epsilon(y - y_c)\delta.
\label{xe}
\end{eqnarray}
Similarly,
\begin{eqnarray}
y - y_e = (1 - \epsilon)(y - y_c) + \epsilon(x - x_c)\delta.
\label{ye}
\end{eqnarray}
By substituting Eqs. (\ref{xe}) and (\ref{ye}) for $x - x_e$ and $y - y_e$ in (\ref{prim}), we get
\begin{eqnarray}
\dot{x} & = & -[(1 - \epsilon) (y - y_c) + \epsilon (x - x_c)\delta]\Omega, \nonumber \\ 
\dot{y} & = & [(1 - \epsilon) (x - x_c) - \epsilon (y - y_c)\delta]\Omega. 
\label{basic_eq}
\end{eqnarray}
These are equations for the polar motion that give the motion of the rotational pole P$(x, y)$
when the location of the figure pole $\textrm{C}(x_c, y_c)$ is given as a known function of time
in the Earth that has elasticity and anelasticity represented by $\mu$ and $\delta$, respectively.
The quantity $\delta$ works so as to damp the amplitude of the Chandler wobble.

Eqs. (\ref{basic_eq}) correspond to the equations that express the motion of the rotational pole around the figure pole that appeared in 
preceding works, Munk and McDonald (1960) and Moritz and Mueller (1987), except that the damping is not taken into account in those works.
In addition, it is not clear whether the rotational pole used in them is the CIP or not.
Further, in Moritz and Mueller (1987), the figure pole, which is the same as the pole C in the present study, is implicitly supposed to be motionless.

The excitation pole in Munk and McDonald (1960) corresponds to the varying figure pole in the present study.
However, the term $``$excitation pole", is unnecessary.  In fact, it is simply the moving figure pole caused by the change in the Earth's physical state.

\section{Polar motion and its primitive analysis}

In this section, we look at the data on polar motion or the coordinates of the rotational pole and carry out a primitive analysis.
The data is taken from the Bulletin B data published by the IERS, which can be downloaded from their website (IERS, 2012).
Of these data, we use the daily data for the period of 20 years (7305 days) starting from 1 January 1992 to 31 December 2011.
It is emphasized that they are all compiled based on the modern observation data acquired by using space-geodetic techniques.

\begin{figure}
\centering
\includegraphics[width = 17cm, clip]{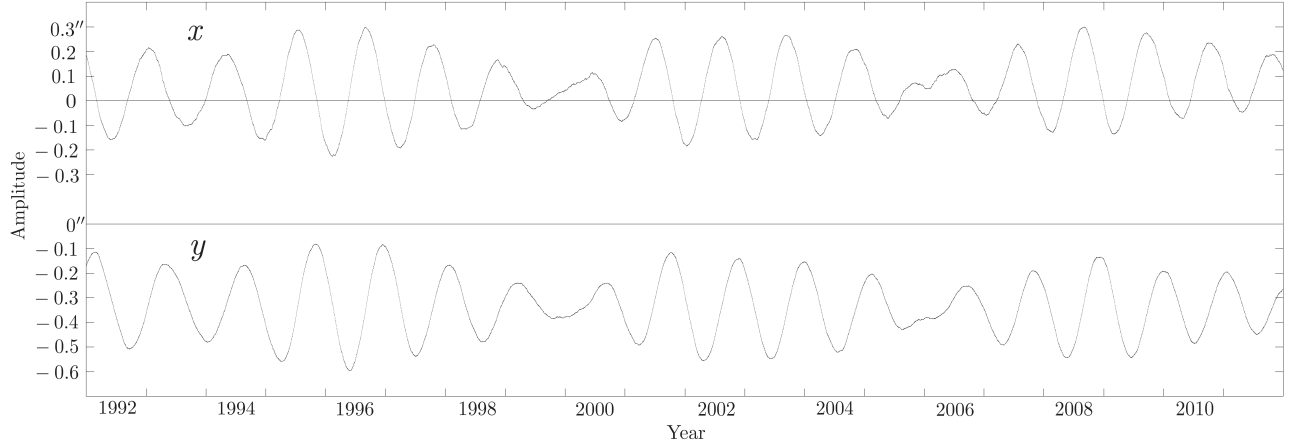}
\caption{Polar motion in the recent 20 years.}
\label{PolMot}
\end{figure}

\begin{figure}
\centering
\includegraphics[width = 8.5cm, clip]{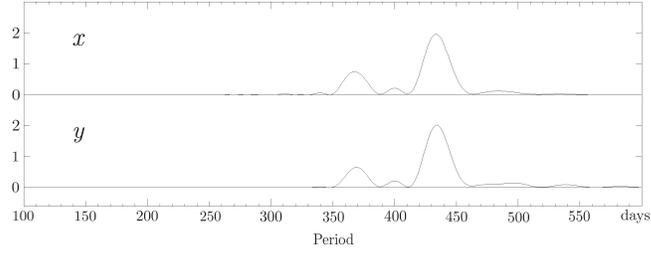}
\caption{Power spectra for the polar motion in the recent 20 years.}
\label{PolMot_F}
\end{figure}

First, we look at the original data of the coordinates of the rotational pole $(x, y)$, which is shown in Fig. \ref{PolMot}.  The origin of the coordinate system is
the International Reference Pole (IRP), and the x- and y-axes are taken in the directions of the Greenwich longitude and $90^{\circ}$ east longitude, respectively.
Therefore, attention should be paid to the sign of y-coordinate, which is opposite to that in the IERS original data.
The power spectra of x- and y-coordinates are shown in Fig. \ref{PolMot_F}.

Polar motion consists of a secular motion and periodic motions.
Of them, we notice first the prevailing period for the Chandler wobble (the free wobble) of about 437 days.
Besides this main component, the annual and semi-annual components are to be recognized.
Overall, all these features are consistent with many preceding analyses,
although some minor differences may exist owing to the difference of the time span submitted for the analysis and so on.

Regarding the coordinates $(x, y)$ shown in Fig. \ref{PolMot}, we express their averaged features as functions of time, fitting a cubic polynomial
to the secular motion, and respective ellipses to the periodic motions of the Chandler wobble and the annual and semi-annual terms,
since these three periods appear clearly in the power spectra in Fig. \ref{PolMot_F}.
That is, we perform the least-squares fitting by assuming the expressions for $x$ and $y$, respectively, as
\begin{eqnarray} 
x & = & a_0 + a_1 t + a_2 t^2 + a_3 t^3\nonumber \\
  &   & + c_1 \sin \Omega't + c_2 \cos \Omega't \nonumber \\
  &   & + c_3 \sin \nu t + c_4 \cos \nu t + c_5 \sin 2 \nu t + c_6 \cos 2 \nu t, \nonumber \\
y & = & b_0 + b_1 t + b_2 t^2 + b_3 t^3\nonumber \\
  &   & + d_1 \sin \Omega't + d_2 \cos \Omega't \nonumber \\
  &   & + d_3 \sin \nu t + d_4 \cos \nu t + d_5 \sin 2 \nu t + d_6 \cos 2 \nu t,
\label{PM_func}
\end{eqnarray}
where $t$ is the time beginning from 1 January 1992, expressed in one tropical year, that is, 365.2422 days.
Also $\Omega' = (2\pi \times 365.2422 / 437)$ radian/yr and $\nu = 2\pi$ radian/yr, where yr denotes the tropical year.

The results are as follows: \\
For the secular motion,
\begin{eqnarray}
a_0 & = & +0.034'' \pm 0.002'', \nonumber \\
a_1 & = & (+0.0022'' \pm0.0008'')/\textrm{yr}, \nonumber \\
a_2 & = & (-0.00036'' \pm0.00009'')/\textrm{yr}^2, \nonumber \\ 
a_3 & = & (+0.000023'' \pm0.000003'')/\textrm{yr}^3, \nonumber \\ 
b_0 & = & -0.318'' \pm 0.002'', \nonumber \\
b_1 & = & (-0.0023'' \pm 0.0008'')/\textrm{yr}, \nonumber \\
b_2 & = & (-0.00013'' \pm 0.00009'')/\textrm{yr}^2, \nonumber \\
b_3 & = & (+0.000010'' \pm 0.000003'')/\textrm{yr}^3.
\label{PM_poly}
\end{eqnarray}
For the Chandler wobble with the period of 437 days,
\begin{eqnarray}
c_1 & = & -0.1001'' \pm 0.0007'', \nonumber \\
c_2 & = & +0.1039'' \pm 0.0007'', \nonumber \\
d_1 & = & +0.1036'' \pm 0.0007'', \nonumber \\ 
d_2 & = & +0.0986'' \pm 0.0007''. 
\label{PM_Ch}
\end{eqnarray}
For the annual variation,
\begin{eqnarray}
c_3 & = & -0.0777'' \pm 0.0007'', \nonumber \\
c_4 & = & -0.0462'' \pm 0.0007'', \nonumber \\
d_3 & = & -0.0424'' \pm 0.0007'', \nonumber \\
d_4 & = & +0.0722'' \pm 0.0007''. 
\label{PM_an}
\end{eqnarray}
For the semi-annual variation,
\begin{eqnarray}
c_5 & = & +0.0043'' \pm 0.0007'', \nonumber \\
c_6 & = & -0.0009'' \pm 0.0007'', \nonumber \\
d_5 & = & +0.0005'' \pm 0.0007'', \nonumber \\
d_6 & = & -0.0014'' \pm 0.0007''. 
\label{PM_s-an}
\end{eqnarray}

The figures after the $``\pm"$ signs are the statistical mean errors.   However, in this case,  they should not be interpreted as the errors
but as the dispersions around the averaged state.

\begin{figure}
\centering
\includegraphics[width = 8.5cm, clip]{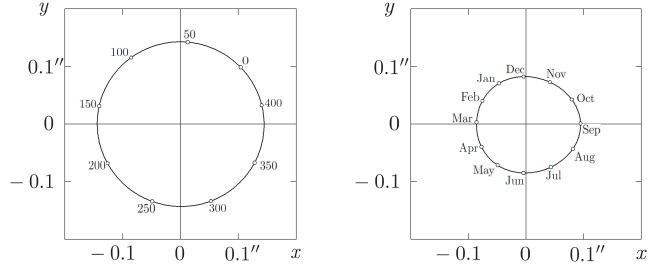}
\caption{Chandler (left) and annual (right) components in the polar motion averaged for 20 years. The semi-annual component is incorporated into the latter.
The numbers in the Chandler component show the days from the start of every period.  The first period begins on 1 January 1992.
The names of the months in the annual component represent the first day of each month.}
\label{PM_comp_S}
\end{figure}%

Eqs. (\ref{PM_Ch}) to (\ref{PM_s-an}) give the loci that are averaged for the duration of 20 years from 1992 to 2011,
of the free wobble and the annual and semi-annual terms and are shown in Fig. \ref{PM_comp_S}, in which the semi-annual component is incorporated into the annual component
since the semi-annual component is very small, when compared to the other.
The two loci exhibit nearly circular patterns.

It should be noticed that, in these figures, the locus for the annual plus semi-annual terms does not represent the varying figure pole in the least.

\section{Determination of the figure pole from polar motion and the estimation of its error}

Now, let us determine the position of the figure axis at any moment from the polar motion data.
From Eqs. (\ref{basic_eq}), we have the following system of equations with respect to $x - x_c$ and $y - y_c$: 
\begin{eqnarray}
& \displaystyle -\epsilon \delta(x - x_c) - (1 - \epsilon)(y - y_c) = \frac{\dot x}{\Omega}, \nonumber \\
& \displaystyle (1 - \epsilon)(x - x_c) - \epsilon \delta(y - y_c ) = \frac{\dot y}{\Omega}.
\end{eqnarray}
By solving these equations, we have
\begin{eqnarray}
x_c & = & x + \frac{\epsilon\delta \dot x - (1 - \epsilon)\dot y}{\Omega[\epsilon^2 \delta^2 + ( 1 - \epsilon)^2]}, \nonumber \\
y_c & = & y + \frac{(1 - \epsilon) \dot x + \epsilon\delta \dot y}{\Omega[\epsilon^2 \delta^2 + ( 1 - \epsilon)^2]}.
\label{fa_sol}
\end{eqnarray}
We notice that these equations are eventually equivalent to those in Wilson (1985) including the damping factor,
although he also uses the term $``$excitation axis" for the rotational axis $(x_c, y_c)$.

From Eqs. \ref{fa_sol} we can obtain the coordinates of the figure pole $(x_c, y_c)$ at any moment.
In order to do this, first we need to give a value for $\delta$.  For this, we assume that $\delta = 1^{\circ}$ tentatively.
We can confirm that the choice of the value does not affect the result much.

Then, we must calculate $\dot x$ and $\dot y$.
We can calculate $\dot x$ by $(x_2 - x_1)/(t_2 -t_1)$, for example, $x_1$ and $x_2$ being the coordinates $x$ at times $t_1$ and $t_2$, respectively.
However, in such a calculation, the accuracy of $\dot x$ and $\dot y$ is feared to be low and we cannot estimate
the error of calculation.  Therefore, we have to calculate $\dot x$ and $\dot y$ in the least-squares method, 
by using the points of P at more points of time.
Thus, we calculate $\dot x$ and $\dot y$ from $2n + 1$ points of P ( $n = 1,2, \dots$), or, in other words, from the data for the time span of
$2n$ days, with the point at the time when its position is to be determined as the center.

As for the error, let us consider evaluating it in the coordinate $x_c$ as an example.
As $\epsilon \cong 0.3$ and $\delta \cong 1^{\circ} \cong 0.017$, we may approximate the first equation of (\ref{fa_sol}) as
\begin{equation}
x_c = x- \frac{\dot y}{\Omega(1 - \epsilon)},
\end{equation}
only for the evaluation of the error.

Let  the errors in $x_c, x$ and $\dot y$ be $\sigma_{x_c}, \sigma_x$ and $\sigma_{\dot y}$, respectively.  Then,
according to  the law for the propagation of errors, we have
\begin{equation}
\sigma_{x_c}^2 = \sigma_x ^2 + \sigma_{\dot y}^2/[\Omega(1 - \epsilon)]^2.
\label{error_law}
\end{equation}  

In this equation, $\sigma_x$ is given in the data by the IERS and it is roughly $ 0.0003''$, which is negligibly small.  Therefore, we only have to calculate $\sigma_{\dot y}$.
For the moment, let the figure pole C be motionless.  As P$(x,y)$ makes a circular motion around C$(x_c, y_c)$ locally,
$\dot y$ is not constant even in this case.  However, we approximate that $y$ is a linear function of time,
letting the error due to this approximation, which is $0.002''$ at the most, be included in the total error.   

\section{Results obtained for the variations in the figure pole}

Now, we examine the results obtained  for the variations in the figure pole as determined by the procedure
described in the previous section.  We examine the results both from the macroscopic and microscopic view-points. 

\  \\
(i) Macroscopic feature of the variation

\begin{figure}
\centering
\includegraphics[width = 17cm, clip]{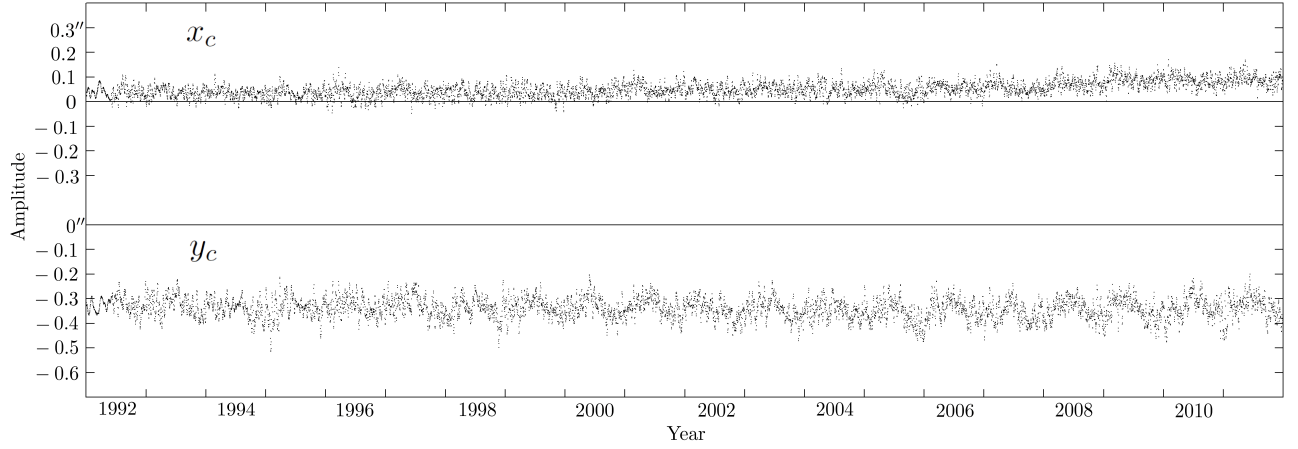}
\caption{Time variation in the coordinates of the figure pole obtained from the polar motion 
shown in Fig. \ref{PolMot}.}
\label{FigPole}
\end{figure}

\begin{figure}
\centering
\includegraphics[width = 8.5cm, clip]{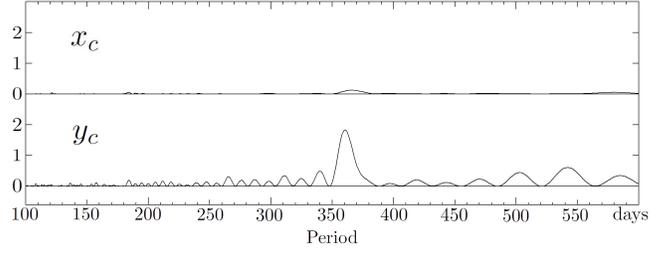}
\caption{Power spectra for the variation in the figure pole shown in Fig. \ref{FigPole}.}
\label{FigPole_F}
\end{figure}

Fig. \ref{FigPole} shows the variation of the coordinates $(x_c, y_c)$ obtained with $n = 1$
and Fig. \ref{FigPole_F} shows the power spectra of the variation.
From Fig. \ref{FigPole_F}, we first notice that the spectra for the variation of the figure pole contain
no frequency for the Chandler wobble of around 437 days, but only the annual and small semi-annual frequencies.

Then, we fit the following equations of time for $x_c$ and $y_c$, which are similar to Eqs. (\ref{PM_func}), but without the terms for the Chandler
wobble, and obtain the coefficients by the least-squares method:
\begin{eqnarray} 
x_c & = & a_0 + a_1 t + a_2 t^2 + a_3 t^3\nonumber \\
  &   & + c_3 \sin \nu t + c_4 \cos \nu t + c_5 \sin 2 \nu t + c_6 \cos 2 \nu t, \nonumber \\
y_c & = & b_0 + b_1 t + b_2 t^2 + b_3 t^3\nonumber \\
  &   & + d_3 \sin \nu t + d_4 \cos \nu t + d_5 \sin 2 \nu t + d_6 \cos 2 \nu t.
\label{FP_func}
\end{eqnarray}

The results are: \\
For the secular motion,
\begin{eqnarray}
a_0 & = & +0.036'' \pm 0.001'', \nonumber \\
a_1 & = & (+0.0001'' \pm0.0005'')/\textrm{yr}, \nonumber \\
a_2 & = & (+0.00001'' \pm0.00006'')/\textrm{yr}^2, \nonumber \\
a_3 & = & (+0.000007'' \pm0.000002'')/\textrm{yr}^3, \nonumber \\
b_0 & = & [-0.331'' \pm 0.002'', \nonumber \\
b_1 & = & (+0.0023'' \pm 0.0008'')/\textrm{yr}, \nonumber \\
b_2 & = & (-0.00056'' \pm 0.00009'')/\textrm{yr}^2, \nonumber \\
b_3 & = & (+0.000022'' \pm 0.000003'')/\textrm{yr}^3.
\label{FP_poly}
\end{eqnarray}
For the annual variation,
\begin{eqnarray}
c_3 & = & +0.0088'' \pm 0.0004'', \nonumber \\
c_4 & = & +0.0013'' \pm 0.0004'', \nonumber \\
d_3 & = & +0.0137'' \pm 0.0006'', \nonumber \\
d_4 & = & -0.0230'' \pm 0.0006''. 
\label{FP_an}
\end{eqnarray}
For the semi-annual variation,
\begin{eqnarray}
c_5 & = & +0.0008'' \pm 0.0004'', \nonumber \\
c_6 & = & -0.0046'' \pm 0.0004'', \nonumber \\
d_5 & = & +0.0026'' \pm 0.0006'', \nonumber \\
d_6 & = & +0.0066'' \pm 0.0006''. 
\label{FP_s-an}
\end{eqnarray}

\begin{figure}
\centering
\includegraphics[width = 3.8cm, clip]{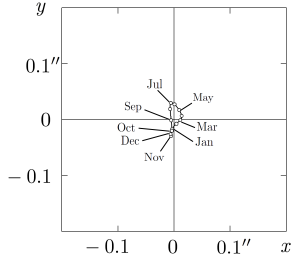}
\caption{Periodic component in the motion of the figure pole on the Earth's surface, averaged for 20 years.
The names of the months represent the first day of each month.}
\label{FP_comp}
\end{figure}%

By examining the above results in reference to those given in Sect. 4,
we can first recognize a non-negligible difference of about $0.01''$ between the constant terms in $(x,y)$ and $(x_c, y_c)$ given by
Eqs. (\ref{PM_poly}) and (\ref{FP_poly}), respectively.
In this connection, we must say that the constant terms of $(x_c, y_c)$ represent a more significant physical entity than those of $(x, y)$.
On the other hand,  it is difficult to say something definite about the secular motions.

Further, when we compare the annual and semi-annual variations in $(x,y)$ and $(x_c, y_c)$, we can observe a striking difference.
The amplitudes of the annual and semi-annual components in $(x_c, y_c)$, given by Eqs. (\ref{FP_an}) and (\ref{FP_s-an}), are much smaller
than those in $(x, y)$ given by Eqs. (\ref{PM_an}) and (\ref{PM_s-an}). 
The locus of the annual plus semi-annual components in the variation of the figure pole is shown in Fig. \ref{FP_comp}.
Of these components, the semi-annual one is very small such that it is almost negligible, and therefore, the locus is eventually that of the annual component alone. 
The locus is far from a circle, but it is rather similar to a line in contrast to that in the polar motion.

\ \\
(ii) Microscopic feature of the variation

\begin{figure}
\centering
\includegraphics[width = 8.5cm, clip]{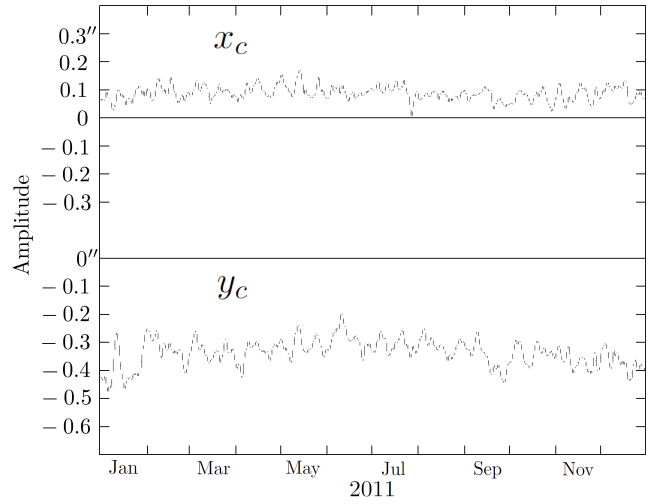}
\caption{Variation in the figure pole in the year 2011 determined by using the values of $\dot x$ and $\dot y$
obtained from the positions of $x$ and $y$ for 3 points of time or the interval of 2 days.
Each point is accompanied by its error bar.}
\label{FP_sh_1}
\end{figure}%

\begin{figure}
\centering
\includegraphics[width = 8.5cm, clip]{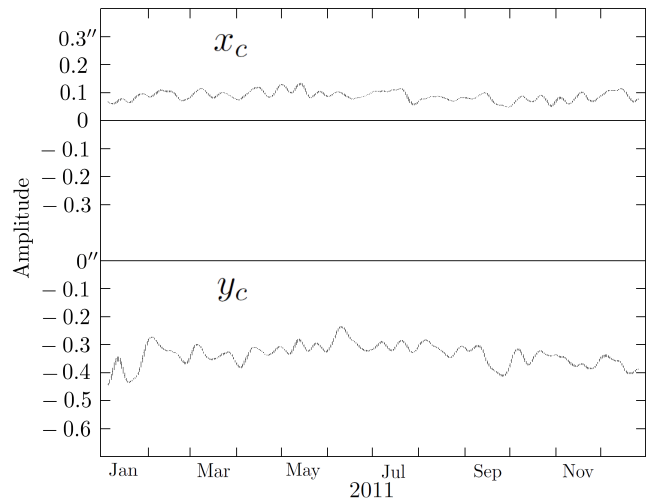}
\caption{Variation in the figure pole in the year 2011 determined by using the values of $\dot x$ and $\dot y$
obtained from the positions of $x$ and $y$ for 11 points of time or the interval of 10 days.
Each point is accompanied by its error bar.}
\label{FP_sh_5}
\end{figure}%

In order to examine the microscopic nature of the variation in the figure pole, 
let us look at the positions and variations in the figure axis as determined in the cases of $n=1$ (Fig. \ref{FP_sh_1}) and $n=5$ (Fig. \ref{FP_sh_5}), respectively,
for a short time interval of one year (the year 2011) but in more detail.
Here, $n$ indicates that $2n + 1$ points of P or the time span of $2n$ days is used for obtaining the results.
The points for $x_c$ and $y_c$ in the figures are accompanied by the mean error bars obtained by the procedure described in the previous section.

From these figures, we see that the values of $\sigma_{x_c}$ and $\sigma_{y_c}$ are not very largely different in both the cases of $n = 1$
(i.e., from 3 points or time span of  2 days) and $n = 5$ (i.e., from 11 points or time span of 10 days), and both are around $0.004''$. 
This situation is almost the same for $n = 2, 3$, and 4.
That is, the error in $(x_c, y_c)$ does not decrease drastically when $n$ is larger.  
% The value of the error $\sigma_{x_c}$ and $\sigma_{y_c}$ are always about $0.004''$.

The reason for this is that the figure pole C moves during
this time span.  The obtained coordinates $(x_c, y_c)$ with a large $n$ are made smooth and dull for some time duration
and therefore, a large number of $n$ is not necessarily desirable.
In other words, the error with $n = 1$ or determined from a time span of 2 days is small enough.
Therefore, the variations that are seen in a small duration of time are significant enough.

When we compare the two graphs given in Figs. \ref{FP_sh_1} and \ref{FP_sh_5}, it is considered that the graph for $n = 1$
has sufficient accuracy with respect to the behavior of the figure axis, while the result for $n >1$ does not add new information, but rather loses
a part of it.

\begin{figure}
\centering
\includegraphics[width = 6cm, clip]{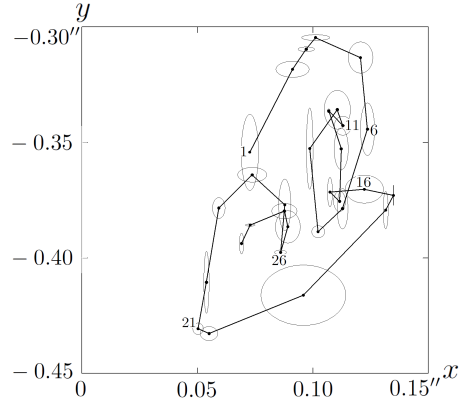}
\caption{Trajectory for the local motion of the figure pole in December, 2011 with the error ellipse at each point.
The numbers represent the days of December, 2011.}
\label{FP_micro}
\end{figure}%

Next, we examine the variation for even a shorter period of one month.
Fig. \ref{FP_micro} is an example for December 2011, which is obtained with $n = 1$.  Each point (day) is accompanied by its error ellipse.
We should recognize that in this rapid motion of the figure pole,
any variation with a magnitude of larger than about $0.004''$ is significant and that it is not a noise,
and that all the small displacements in the motion faithfully reflect the changes in the Earth's state.   

\section{Conclusions}

Since polar motion is the motion of the rotational pole on the Earth's surface,
we can detect the motion of the figure pole relative to the Earth's surface if we know the relation between the
rotational pole and the figure pole.  The variation in the figure pole obtained in this way reflects
the physical state of the Earth directly, and therefore, it is a powerful means to diagnose it.

In the present study, we first derived the equations to determine the position of the figure pole and, by using the data on
polar motion published by the IERS for the duration of 1 January 1992 to 31 December 2011,
we determined the position of the figure pole for every day in this period.

From a macroscopic view-point with a long range of time span, the averaged motion of the figure pole has an obvious annual component
with an amplitude of about $0.02''$, while it has no component of the Chandler period.
It is noticeable that the annual component (almost the averaged motion itself) of the figure pole is quite different from that contained in the polar motion. 

The estimated error of the determination is about $0.004''$ and therefore, the obtained microscopic variation
in the figure pole for a very short time span, such as even one day, is significant enough and it faithfully reflects the Earth's state.

\section{Acknowledgment}
The author is grateful to Prof. Hiroshi Kinoshita for his valuable comments.

\end{document}